          [2001/04/25 1.1 (PWD)]
\documentclass[a4paper,twocolumn]{esapub} 
\usepackage{natbib}
\usepackage{graphicx}

\title{RV survey for planets of brown dwarfs and~very~low-mass~stars~in~Cha\,I}
\author[1]{Viki Joergens}
\affil[1]{Max-Planck-Institut f\"ur extraterrestrische Physik, 
Giessenbachstr. 1, D-85748 Garching, Germany}
\author[1,2]{Ralph Neuh\"auser}
\affil[2]{Astrophysikalisches Institut der Universit\"at von Jena,
		Schillerg\"asschen 2-3, D-07745 Jena, Germany}

\begin{document}

\keywords{extrasolar planets, brown dwarfs, planet formation,
brown dwarf formation}

\maketitle

\begin{abstract}
We have carried out a radial velocity (RV) search for planets and 
brown dwarf companions to 
very young (1-10\,Myr) brown dwarfs and
very low-mass stars in the Cha\,I star forming region.
This survey has been carried out with the 
high-resolution Echelle spectrograph UVES at the VLT.
It is sensitive down to Jupiter mass planets.

Out of the twelve monitored very low-mass stars and brown dwarfs, ten have
constant RVs in the presented RV survey. This hints at a 
small multiplicity fraction 
of the studied population of brown dwarfs and 
very low-mass stars in Cha\,I at small separations.
Upper limits for the mass M$_2 \sin i$ of possible companions have been  
estimated to range between 0.1\,M$_\mathrm{Jup}$ and 1.5\,M$_\mathrm{Jup}$. 
However, two very low-mass stars in Cha\,I
show significant radial velocity variations.
The nature of these variations is still unclear. If caused by orbiting
objects the recorded variability amplitudes would correspond to 
planets of the order of a few Jupiter masses.
%

Furthermore, as a by-product of the RV survey for companions, 
we have studied the kinematics of the brown dwarfs
in Cha\,I. Precise kinematic studies of young brown dwarfs are interesting
in the context of the question if brown dwarfs are
formed by the recently proposed ejection scenario. 
We have found that the RV dispersion of brown dwarfs in Cha\,I 
is only 2.2\,km\,s$^{-1}$
giving a first empirical upper limit for possible 
ejection velocities.

\end{abstract}

\section{Introduction}

Most of the extrasolar planets known to date are considerably old,
with ages of the order of a few billion years.
The youngest planet known to date is orbiting
the zero-age main sequence star $\iota$\,Hor, with an estimated age in 
the range of 30\,Myr to 2\,Gyr (K\"urster et al. 2000).
Furthermore, 
most planets known to date orbit around solar-mass stars with spectral 
types of 
late-F, G and early-K with the exception of
a planet around the M4-dwarf Gl\,876, which has an estimated mass of
0.3--0.4\,M$_{\odot}$ (Delfosse et al. 1998, Marcy et al. 1998).

The search for planets around very young as well as around 
very low-mass stars 
(and even down to the substellar regime)
is particularly interesting since 
the detection of \emph{young} planets as well as a census of planets around 
stars of all spectral types, and maybe even around brown dwarfs,
is an important step
towards the understanding of planet formation. 
It would provide empirical 
constraints for planet formation time scales. 
Furthermore, it would show if planets can exist around objects 
which are of considerably lower mass and surface 
temperature than our sun. 

We have therefore carried out a RV survey
for (planetary and brown dwarf) companions to young brown 
dwarfs and very low-mass stars in the Cha\,I star forming region.
It is aimed at the finding of young planets and brown dwarf
companions with an age 
of a few million years around mid- to late-M dwarfs
close or below the Hydrogen burning limit.

The detection of planets around brown dwarfs as well as the detection of
young spectroscopic brown dwarf binaries (brown dwarf--brown dwarf pairs)
would, on the other hand, be an important clue towards the formation of 
brown dwarfs. 
So far, no planet is known orbiting a brown dwarf.
There have been detected several brown dwarf binaries, among them
there are three spectroscopic, and hence close,
brown dwarf binaries (Basri \& Mart\'{\i}n 1999,
Guenther \& Wuchterl 2003). 
However, all known brown dwarf binaries are fairly old and 
it is not yet established if the typical outcome of 
the brown dwarf formation process is a binary or multiple brown dwarf system
or a single brown dwarf. Furthermore it is not known, if 
brown dwarfs can have planets.

\section{Sample}

The targets of our RV survey are a population of 12 young 
bona fide and candidate brown dwarfs (spectral type M6--M8)
in the center of the Cha\,I star forming cloud (distance $\sim$160\,pc)
with ages of 1--5\,Myr.
They have been detected by Comer\'on et al. (1999, 2000) and
Neuh\"auser \& Comer\'on (1999) and are named Cha\,H$\alpha$1--12.
Furthermore, our sample includes several young ($<$ 10\,Myr),
very low-mass T~Tauri stars in the region, B\,34,
CHXR\,74 and Sz\,23, which have spectral types of M2.5--M5 and
masses of about 0.1--0.3\,M$_{\odot}$ (Comer\'on et al. 1999). 

\section{UVES spectra and RV~measurements}

High-resolution ($\lambda / \Delta \lambda$ = 40\,000) spectroscopy
has been performed 
of the objects in the red region of the optical 
wavelength range (6700\,{\AA}--1\,$\mu$m) with the Echelle spectrograph
UVES at the 8.2\,m VLT Kuyen telescope at ESO Paranal, Chile.
We took several spectra, at least two, of each of the objects in 2000 and 2002.
We have measured RVs with a precision of down to 40\,m/s 
by means of cross-correlating plenty of stellar lines of the 
object spectra with a template spectrum
using telluric lines as wavelength reference
(Joergens \& Guenther 2001; Joergens 2003).
The RV precision of the relative RVs ranges between 40\,m/s and 600\,m/s
depending on the S/N of the individual spectra. The RV errors have been 
estimated by 
the deviations of two consecutive RV measurements based on two single consecutive
spectra.

\section{Results}

\subsection{Low multiplicity of brown dwarfs}

\begin{figure}
\centering
\includegraphics[width=0.9\linewidth]{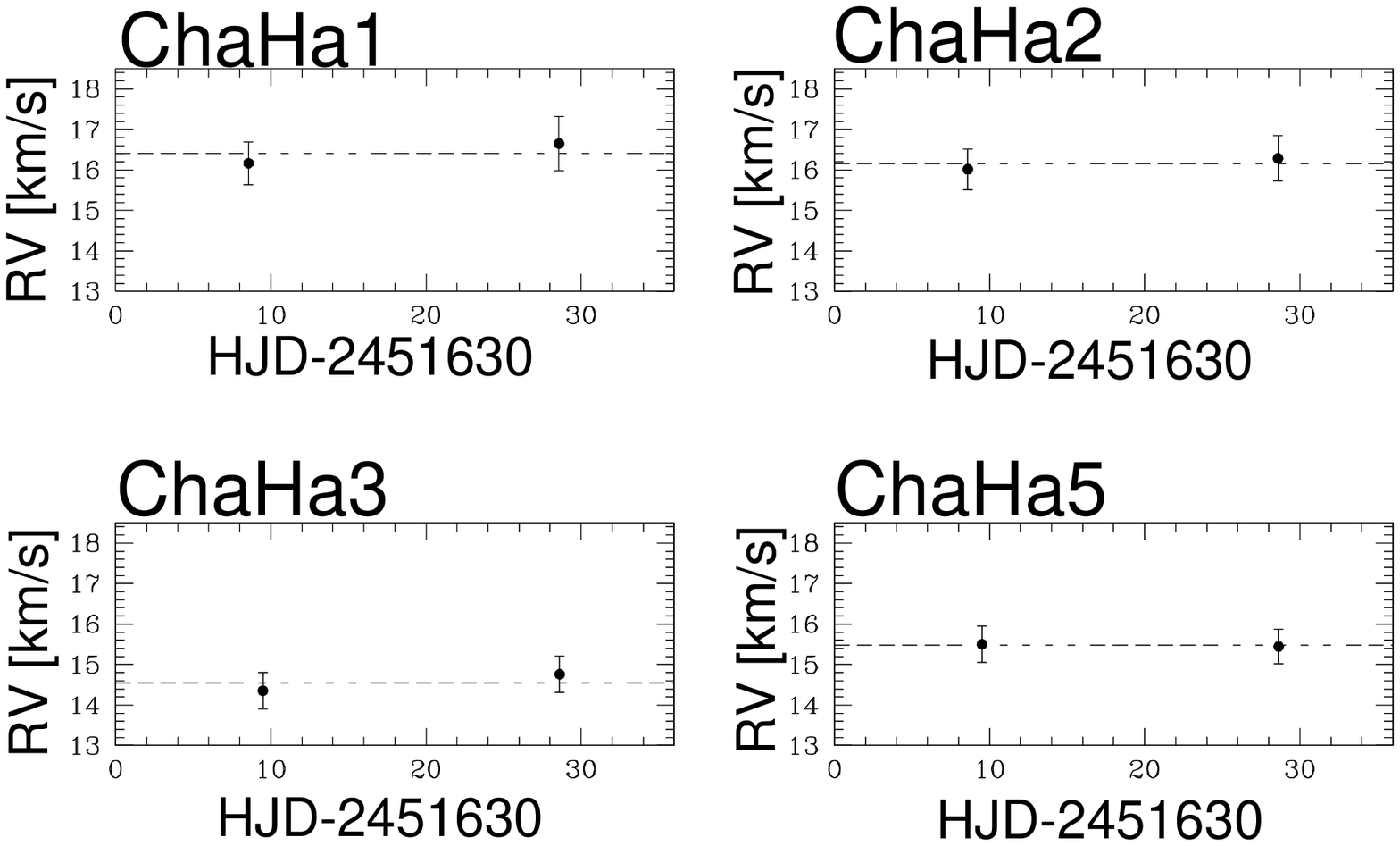}
\includegraphics[width=0.9\linewidth]{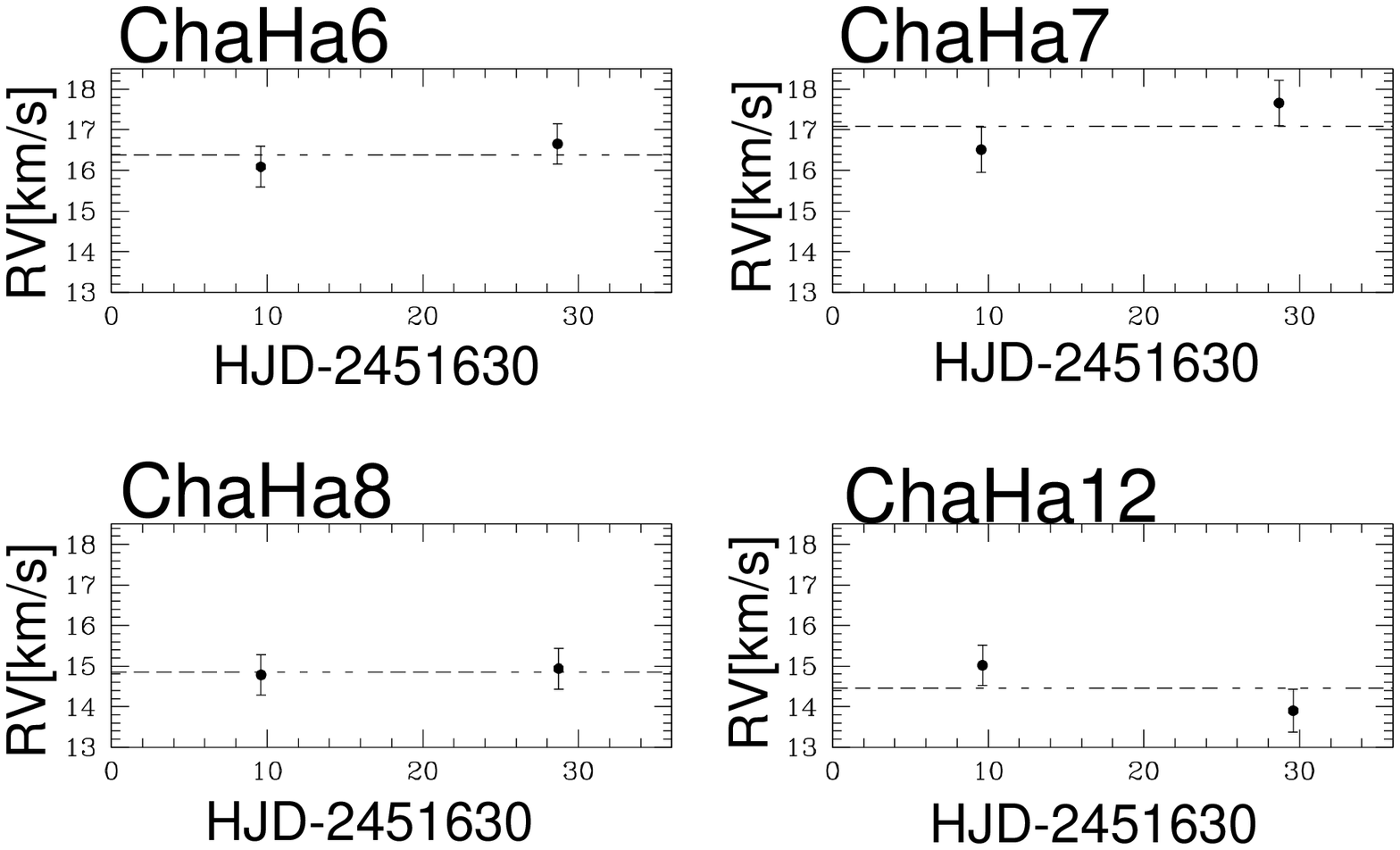}
\caption{
\label{bds}
RV vs. time in Julian days for bona fide and candidate brown dwarfs in Cha\,I
based on high-resolution UVES/VLT spectra.
Error bars indicate 1 $\sigma$ errors.
The objects show no RV variability.
}
\end{figure}

\begin{figure}
\centering
\includegraphics[width=0.9\linewidth]{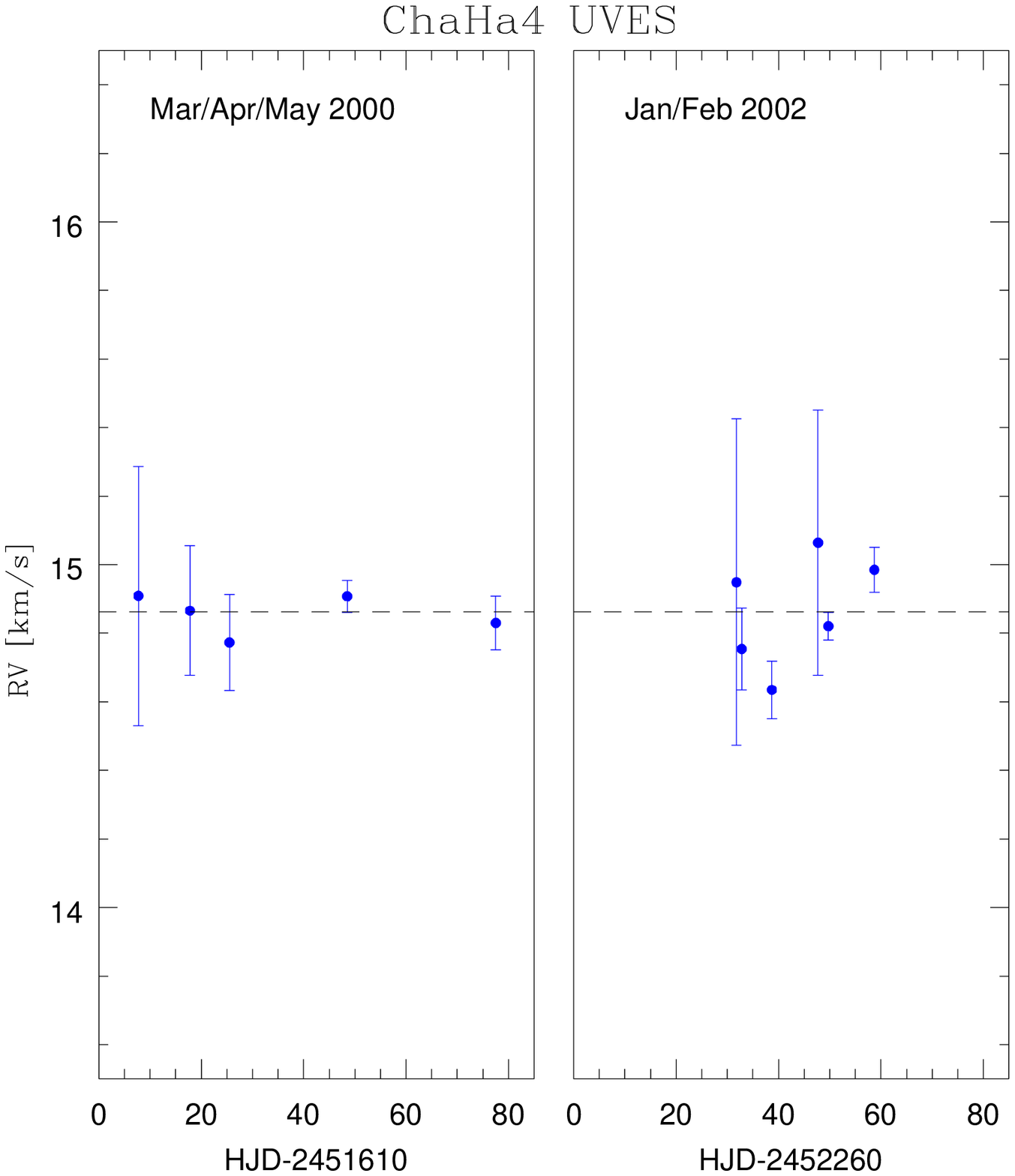}
\caption{
\label{cha4}
RV vs. time in Julian days for the candidate brown dwarf Cha\,H$\alpha$\,4
based on high-resolution UVES/VLT spectra.
Error bars indicate 1 $\sigma$ errors.
The object shows small RV variability but not exceeding 2\,$\sigma$.
}
\end{figure}

We have found that the radial velocities for 
the bona fide and candidate brown dwarfs
Cha\,H$\alpha$\,1, 2, 3, 5, 6, 7, 8 and 12
are constant within the measurements uncertainties (Fig.\,\ref{bds}).
Cha\,H$\alpha$\,4 shows small variations (Fig.\,\ref{cha4}), 
however, they do not exceed 2\,$\sigma$ (Joergens et al. 2002, 
Joergens 2003).

We have estimated upper limits for masses of hypothetical companions 
$M_2 \sin i$
of these objects by assuming that the total variability amplitude was 
recorded. 
They range between 0.3\,M$_\mathrm{Jup}$ and 1.5\,M$_\mathrm{Jup}$
based on the assumption 
of a circular orbit and a separation of 0.1\,AU between companion and 
primary object.
Primary masses have been taken from Comer\'on et al. (1999, 2000).

That means, that no planet around a young brown dwarf has been found 
and, furthermore, also
the multiplicity fraction for close binary brown dwarfs among the sample
is apparently rather low. 
There is, of course, 
the possibility that present companions have not been detected 
due to non-observations at the critical orbital phases.
Furthermore, long-period companions may have been missed.
However, the found small multiplicity fraction of the bona fide and candidate 
brown dwarfs in Cha\,I at small separations,
is also supported by the result of a direct imaging search for wide 
(planetary or brown dwarf) companions 
to the same targets, Cha\,H$\alpha$\,1--12, by 
Neuh\"auser et al. (2002a, 2002b), who find a
multiplicity fraction of $\leq$10\%.

\subsection{Young Jupiter mass planets ?}

\begin{figure}
\centering
\includegraphics[width=0.9\linewidth]{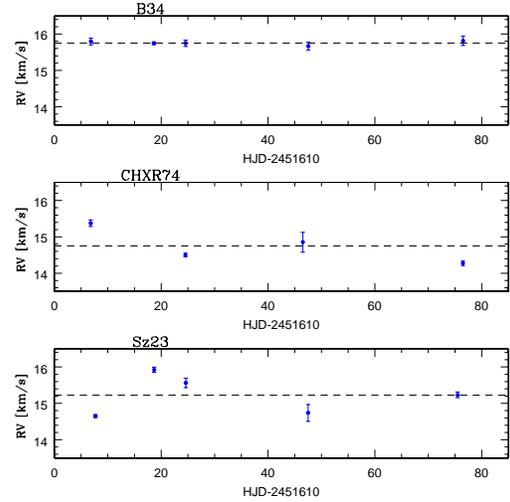}
\caption{
\label{tts}
RV vs. time in Julian days for the very low-mass T~Tauri stars
B\,34, CHXR\,74 and Sz\,23
based on high-resolution UVES/VLT spectra.
Error bars indicate 1\,$\sigma$ errors.
The RVs of CHXR\,74 and Sz\,23 are significantly variable, 
whereas B\,34 shows no variability.
}
\end{figure}

The RV survey of the very low-mass T~Tauri stars in Cha\,I
showed that the RVs of B\,34 
are constant within the measurements uncertainties, 
as shown in the top panel of Fig.\,\ref{tts}. 
Furthermore, it revealed significant RV variations for
CHXR\,74 ($\sim$0.17\,M$_{\odot}$) and Sz\,23 ($\sim$0.3\,M$_{\odot}$); 
cf. middle and bottom panel of Fig.\,\ref{tts}.

\begin{figure}
\centering
\includegraphics[width=0.9\linewidth]{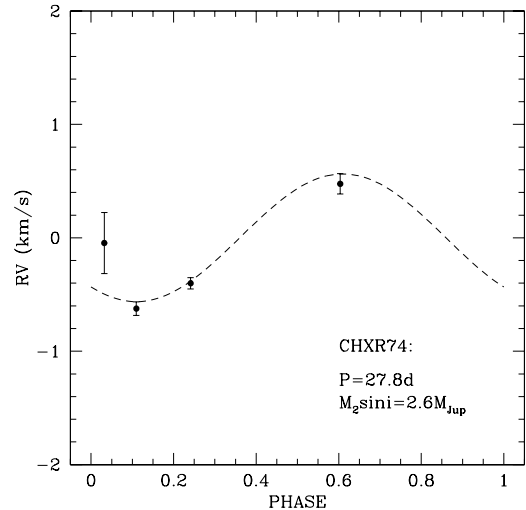}
\caption{
\label{orbit}
Hypothetical RV curve for CHXR\,74 based on the RV data shown 
in Fig.\,\ref{tts}
for a planetary companion with mass M$_2 \sin i$ of 
2.6\,M$_{\mathrm{Jup}}$ in a circular 28 day orbit. 
The first data point is off by less than 1.5\,$\sigma$.
However, follow-up observations have to explore 
the nature of the detected RV variatons.
}
\end{figure}

The nature of these RV variations is still unclear.
They could be caused by surface activity since prominent surface spots
can cause RV variability due to the shifting of the photocenter
while the star rotates.
The other possibility is that they are caused by an orbiting mass
pulling on the primary and causing a Doppler wobble.
If caused by orbiting companions, the detected RV variations
correspond to planets of a few Jupiter masses in close orbits 
around CHXR\,74 and Sz\,23.
As an example, 
a hypothetical orbit has been found for CHXR\,74, 
which matches the recorded RV data (Fig.\,\ref{orbit}). It 
shows that the RVs are consistent with a companion mass M$_2 \sin i$
of 2.6\,M$_{\mathrm{Jup}}$ orbiting in a circular orbit with a period of
28\,d.
The upper limits for the rotational periods of CHXR\,74 and Sz\,23
are of the order of a few days based on projected rotational
velocities $v \sin i$ (Joergens \& Guenther 2001), 
i.e. a 28\,d periodicity, for example, cannot be 
caused by rotational modulation due to surface features.

In order to explore the nature of the detected 
RV variations and, if caused by companions, to solve 
the spectroscopic orbit,
follow-up observations of CHXR\,74 and Sz\,23 are planned.
If confirmed as planetary systems, they
would be unique because they would contain not only the lowest mass primaries
but with an age of a few million years 
also the by far youngest extrasolar planets found to date giving first
empirical constraints for planet formation time scales.

\subsection{Formation of brown dwarfs: 
kinematics of brown dwarfs in Cha\,I}

Brown dwarfs are an important link between the two distinct
populations of planets and stars.
We still do not know by which mechanism brown dwarfs form. 
However, we expect that the exploration of 
the formation mechanism producing brown dwarfs may shed light also on 
some details of the planet formation.
Recently, it has been proposed
that brown dwarfs are formed due to the ejection of protostars in the 
early accretion phase, which are thenceforward cut off from the gas reservoir
and cannot accrete to stellar masses (Reipurth \& Clarke 2001).

Based on mean RVs derived within the presented RV survey, 
we have carried out a precise kinematic study 
of the bona fide and candidate brown dwarfs in Cha\,I.
It showed that the RV dispersion of the brown dwarfs 
is relatively small, 2.2\,km\,s$^{-1}$, and that the RVs cover a total range 
of only 2.6\,km\,s$^{-1}$, indicating that 
none of the studied brown dwarfs has been ejected with higher velocity
out of their birth place (Joergens \& Guenther 2001).
The RV dispersion of the brown dwarfs is significantly 
smaller than that of the 
T~Tauri stars in the same field (3.6\,km\,s$^{-1}$) and slightly larger 
than that of the surrounding molecular gas (1.2\,km\,s$^{-1}$).

Very recent dynamical calculations
(Sterzik \& Durisen 2002; Bate et al. 2003)
yield rather small ejection velocities for brown dwarfs, 
suggesting the possibility
that the imprint of the ejection in the kinematics might not be an observable 
effect.
However, these theoretical values for ejection velocities of brown dwarfs
rely on certain model assumptions and are still a matter of debate.
On the other side, 
the kinematic study presented by us contributes a first 
observational constraint for the velocity distribution 
of a homogenous group of closely confined very young brown 
dwarfs and therefore an \emph{empirical} upper limit for possible ejection 
velocities.

\section*{Acknowledgments}
We are very grateful to Eike Guenther for significant 
help with the spectroscopic observations and data analysis.
Furthermore, we also like to thank Fernando Comer\'on and 
Matilde Fernand\'ez for fruitful collaborations.

\end{document}